\documentclass[twocolumn,times,trackchanges, dvipsnames]{aastex631}
\usepackage[utf8]{inputenc}
\usepackage{amsmath}

\begin{document}
\title{Bolometric Corrections for FU Ori Object Accretion Disk Luminosities}

\author{Adolfo S. Carvalho}
\affiliation{Department of Astronomy; California Institute of Technology; Pasadena, CA 91125, USA}
\author{Lynne A. Hillenbrand}
\affiliation{Department of Astronomy; California Institute of Technology; Pasadena, CA 91125, USA}

\begin{abstract}
    The accretion luminosity of an FU Ori disk is a fundamental system parameter, but a challenging one to estimate for all but the most well-studied systems. FU Ori objects are dynamically evolving accretion disks, especially close in time to the outburst epoch. They have a complex multi-temperature disk structure that results in distinctly shaped, broad SEDs. Detailed spectroscopic analysis is required for simultaneous constraint on relevant physical parameters such as the central stellar mass, inner disk radius, disk inclination, and disk accretion rate. However, outbursting systems that are deeply embedded and/or distant may be limited to only photometric measurement, and over only a narrow range of wavelengths. The bolometric corrections necessary to estimate accretion luminosities are not straightforward, and in particular can not be adopted from existing literature on isotropically radiating stars. We present bolometric corrections specific to astrophysical accretion disks for a variety of filters in ongoing and upcoming all-sky surveys.
\end{abstract}


\section{Introduction}\label{sec:introduction}
FU Ori outbursts occur when the accretion disks around young stellar objects (YSOs) undergo instabilities that increase their accretion rates by factors of $10^2-10^4$ \citep{hartmann_fu_1996}. The outbursts have been observed to last for decades and some objects, like V883 Ori, are estimated to have been in outburst for over 100 years \citep{StromStrom_V883OriDiscovery_1993ApJ}. When YSOs are in this luminous, long-lived, outbursting state they are labeled as FU Ori objects. 

Extreme accretion outbursts like those of FU Ori objects are proposed as a solution to the ``luminosity problem", a timescale problem in stellar mass assembly \citep{Kenyon_TaurusLuminosityProblem_1990AJ, kenyon_hartmann_1995, Myers_LuminosityProblemSmaller_2010ApJ, fischer_stellarMassAssembly_2023ASPC}. 
The problem arises from the observation that measured mass accretion rates of quiescent Class 0 to Class II YSOs are much too low for the forming stars to acquire their eventual main sequence mass in the estimated lifetimes of the envelope (0.5 Myr) and disk ($<$5-10 Myr); \cite{Evans_LowMassStarFormationReview_2011IAUS}. 

One of the greatest limitations to better understanding the FU Ori phenomenon is the small sample size, as fewer than 45 outbursts have been observed \citep{connelley_near-infrared_2018, Guo_VVV_FUOriBursts_2024MNRAS}. Fortunately, the number of FU Ori objects discovered each year has increased exponentially 
as more all sky surveys have turned on at different wavelengths (Hillenbrand 2025, in prep). While many of the early FU Ori outbursts  (FU Ori, V1057 Cyg, V1515 Cyg) were discovered by astute professional astronomers, a large number of them in the 1980s-2000's were found by dedicated citizen scientists. Most recently, the PTF/ZTF \citep{Law_PTFReference_2009PASP, Bellm_ZTFReference_2019PASP}, Gaia \citep{GaiaMissionReference_2016A&A}, Gattini \citep{Moore_gattini_2016SPIE, De_PalomarGattiniIRReference_2020PASP}, VVV/VVVX \citep{Minniti_VVV_2010NewA, Saito_VVVX_2024A&A}, and NEOWISE \citep{Mainzer_neowise_2011ApJ} time series photometry surveys have resulted in the detection of tens of FU Ori outbursts over the past 15 years, more than doubling the sample of observed outbursts. 

In the coming years, two major all sky surveys will begin: the Rubin Observatory Legacy Survey of Space and Time (LSST), in the visible, and the Roman Observatory High Latitude Wide Area Survey and Galactic Microlensing Survey, in the infrared. The increased depth of these surveys relative to existing ones is expected to enable the observation of new FU Ori eruptions.

However, many of these objects will be distant, and likely at least partially embedded. They will be thus difficult to characterize in detail. In particular, it may be impossible with current capabilities to assemble comprehensive SEDs or obtain even limited wavelength multi-epoch spectroscopy.  For many, perhaps most, of the photometrically detected outbursts, only distinguishing lightcurves will be available.

Here, we provide a series of bolometric corrections for FU Ori objects in common photometric filters. We include those from both existing (ZTF, Gaia, 2MASS, NEOWISE) and upcoming (LSST, Roman) all sky time domain surveys. With the proper corrections for accretion disks from observed to bolometric magnitudes, and with further correction for distance and extinction to absolute bolometric magnitudes, the accretion luminosities (and thence accretion rates) of FU Ori objects can be constrained .

\section{Applying A Full-Spectrum Disk Model to Photometric Observations} \label{sec:SEDModel}
\subsection{SED for a typical FU Ori object}
We construct a model SED of an FU Ori object assuming that in the visible/near-infrared the emission is dominated by a viscously heated accretion disk. Our model follows the method put forth by \citet{Kenyon_FUOri_disks_1988ApJ} but includes additional elements, as
described in detail in \citet{Carvalho_V960MonPhotometry_2023ApJ}.  We summarize the model here. 

First, we assume the accretion disk is thin and viscously heated. We then adopt a modified version of the \citet{Shakura_sunyaev_alpha_1973A&A} $\alpha$-disk temperature profile, 
\begin{equation} \label{eq:TProf}
    T^4_\mathrm{eff}(r) = \frac{3 G M_* \dot{M}}{8 \pi \sigma r^3} \left( 1 - \sqrt{\frac{R_\mathrm{inner}}{r}}  \right)    ,
\end{equation}
where $R_\mathrm{inner}$ is the inner radius of the accretion disk, $M_*$ is the mass of the central star, $\dot{M}$ is the stellar accretion rate, $G$ is the gravitational constant, and $\sigma$ is the Stefan-Boltzmann constant. We assume that $T_\mathrm{eff}(r < \frac{49}{36} R_\mathrm{inner}) = T_\mathrm{eff}(\frac{49}{36}R_\mathrm{inner}) = T_\mathrm{max}$ \citep{Kenyon_FUOri_disks_1988ApJ}. We then populate the annuli of the disk with PHOENIX model spectra \citep{Husser_Phoenix_2013A&A} with log$g=1.5$ and the appropriate $T_\mathrm{eff}(r)$. 

For the physical parameters of the star and disk, we adopt those of FU Ori \citep{Zhu_outburst_FUOri_2020MNRAS}: $M_* = 0.6 \ M_\odot$, $R_\mathrm{inner} = 3.52 \ R_\odot$, $R_\mathrm{outer} = 0.7$ AU. Since we will be computing bolometric corrections for various system luminosities, we vary $\dot{M}$ in the range $10^{-5.5}-10^{-3.5}$ $M_\odot$ yr$^{-1}$ to cover $8 \ L_\odot < L_\mathrm{acc} < 845 \ L_\odot$ and $3,300 \ \mathrm{K} < T_\mathrm{max} < 10,500$ K. The grid of luminosities and effective temperatures slightly exceeds the range of values found in detailed modelling of FU Ori objects \citep{Nayakshin_TI_FUOriOutbursts_2024MNRAS}.

The model grid is computed for systems at distance $d = 10$ pc, with extinction $A_V$ = 0 mag. We also assume the disk is face on (inclination $i = 0^\circ$). Below, when applying the model grid to observations, the latter are corrected to these parameter values.

\subsection{The filter profiles}

We obtained filter profiles and photometric zero points from the Spanish Virtual Observatory (SVO) Filter Profile Service\footnote{\url{http://svo2.cab.inta-csic.es/svo/theory/fps3/index.php?mode=browse}}. We selected several major large-area time domain surveys to represent: the on-going ZTF and Gaia,  and the upcoming LSST optical surveys, the upcoming Roman near-infrared survey, and the recently completed NEOWISE mid-infrared survey that probes the most embedded YSOs.
We additionally include the 2MASS $J$, $H$, and $Ks$ bands due to the broad use of 2MASS as a calibration for other $J$, $H$, and $K$ filter systems in various time domain surveys \citep[e.g., Gattini-IR and WINTER:][]{Moore_gattini_2016SPIE, lourie_winter_instrument_2020SPIE11447E}.

\begin{figure}
    \centering
    \includegraphics[width = \linewidth]{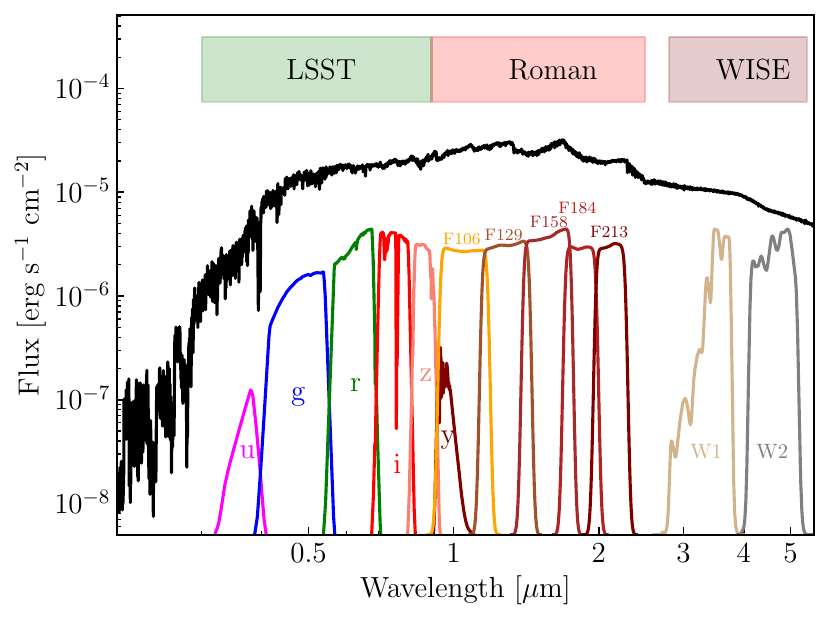}
    \caption{The normalized filter profiles for the LSST, Roman, and WISE filters (colored lines), plotted against our fiducial $\dot{M} = 10^{-4.5} \ M_\odot$ yr$^{-1}$ FU Ori object SED model (black). The filled boxes at the top show the approximate total wavelength coverage of each survey or mission. The optical ZTF and Gaia filters overlap the LSST bands, and the 2MASS filters overlap the Roman bands.
    }
    \label{fig:FilterProfiles}
\end{figure}

The filter profiles for LSST, Roman, and WISE are shown in Figure \ref{fig:FilterProfiles} for reference. Notice that the bolometric corrections will be largest for the $u$, $g$, $W1$, and $W2$ filters, while the SED of a typical FU Ori disk is relatively flat for $0.4 < \lambda < 1.3 \ \mu$m. 

\subsection{Bolometric Corrections} \label{sec:bolCorr}

\begin{figure*}
    \centering
    \includegraphics[width = 0.32\linewidth]{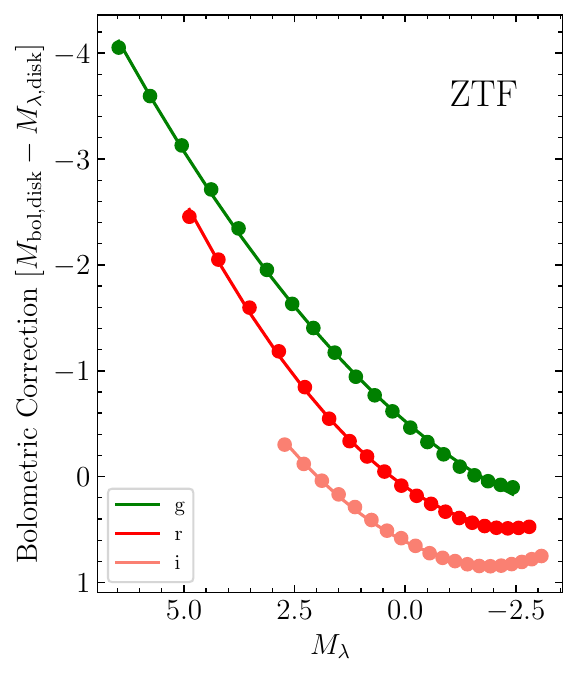}
    \includegraphics[width = 0.32\linewidth]{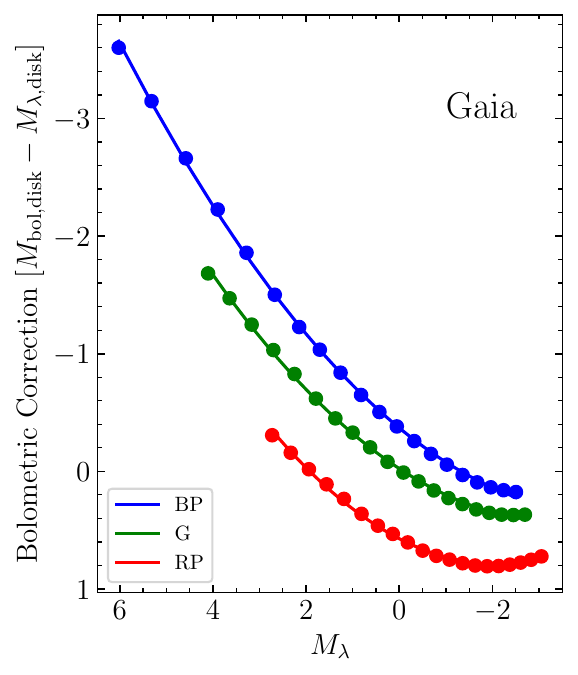}
    \includegraphics[width = 0.32\linewidth]{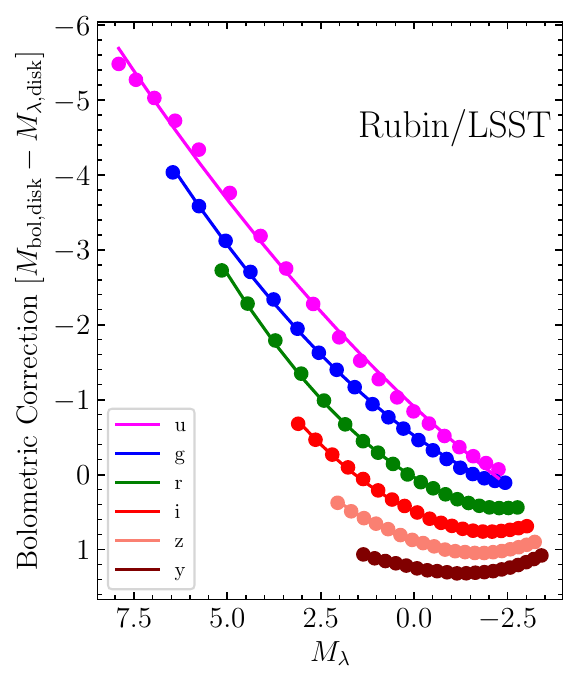} \\

    \includegraphics[width = 0.32\linewidth]{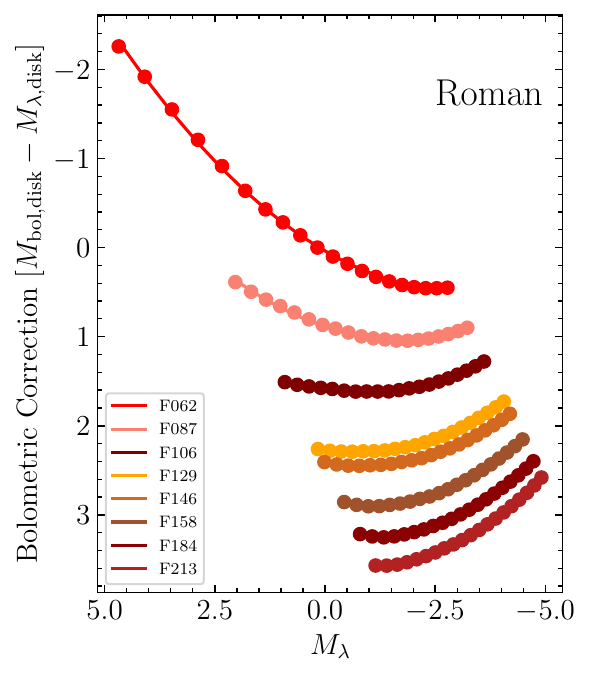}
    \includegraphics[width = 0.32\linewidth]{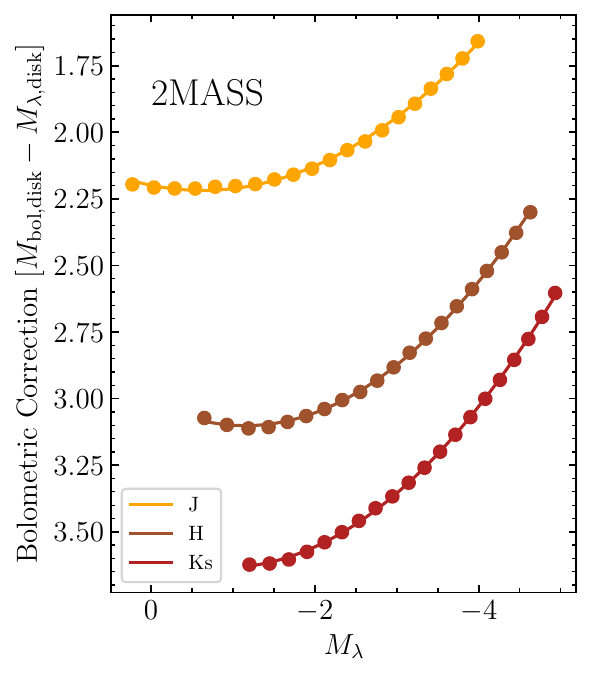}
    \includegraphics[width = 0.32\linewidth]{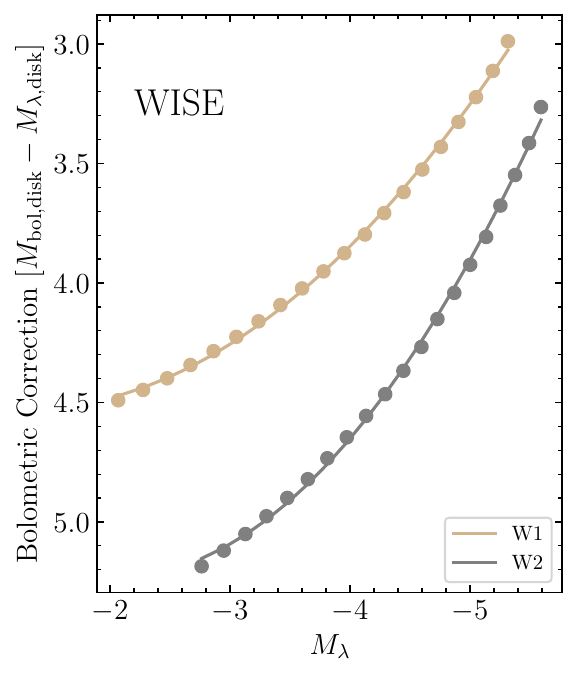}
    \caption{The bolometric corrections (filled circles) and our polynomial fits (solid lines). The visible range surveys in the top row are ZTF (left), Gaia (center), and LSST (right). The infrared surveys along the bottom are Roman (left), 2MASS (center), and WISE (right).  }
    \label{fig:BCs}
\end{figure*}

For each of the filters, we compute the observed photometric magnitude in our model, which is by construction the absolute magnitude. We then compute the bolometric correction, $BC_{\lambda,\mathrm{disk}} \equiv M_\mathrm{bol,disk} - M_{\lambda,\mathrm{disk}}$ between the absolute magnitude in the observed band and the bolometric magnitude, assuming a solar bolometric magnitude of $M_{\mathrm{bol}, \odot} = 4.74$ \citep{Willmer_AbsMagSun_2018ApJS}. The bolometric corrections are shown as a function of absolute magnitude, $M_{\lambda,\mathrm{disk}}$, in Figure \ref{fig:BCs}.

We can fit the $BC_{\lambda,\mathrm{disk}}$ versus $M_{\lambda,disk}$ relation with a 2nd order polynomial:
\begin{equation} \label{eq:Bcorr}
    BC_{\lambda,\mathrm{disk}} = c_0 + c_1 M_{\lambda,\mathrm{disk}} + c_2 M_{\lambda,\mathrm{disk}}^2,
\end{equation}
where $BC_{\lambda,\mathrm{disk}}$ is the bolometric correction for a given filter, $\lambda$, and $c_0$, $c_1$, $c_2$ are the fit coefficients. The best-fit polynomials are also plotted in Figure \ref{fig:BCs}. We report the coefficients and the reference wavelength of each filter in Table \ref{tab:BC_tab}.

\begin{deluxetable}{c|c|ccc|c}
	\tablecaption{Polynomial coefficients $c_0,c_1,c_2$ in front of absolute magnitudes (see Equation \ref{eq:Bcorr}) for the bolometric corrections of FU Ori objects. Also provided are $A_{\lambda,\mathrm{disk}}/A_V$ values specifically computed for the viscous disk model. 
 \label{tab:BC_tab}}
	\tablewidth{0pt}
	\tablehead{
	    \colhead{Filter} & \colhead{$\lambda_\mathrm{ref}$ $\left[ \mathrm{\AA} \right]$ } & \colhead{$c_0$} & \colhead{ $c_1$ } & \colhead{$c_2$} & \colhead{$A_{\lambda, disk} / A_V$}
	}
\startdata
Gaia BP & 5041.61 & -0.368 & -0.326 & -0.036 & 0.84 \\
Gaia G & 5850.88 & -0.022 & -0.260 & -0.038 & 0.50 \\
Gaia RP & 7690.74 & 0.574 & -0.205 & -0.048 & 0.46 \\
\hline
ZTF g & 4746.48 & -0.522 & -0.357 & -0.030 & 1.07 \\
ZTF r & 6366.38 & 0.097 & -0.289 & -0.051 & 0.72 \\
ZTF i & 7829.03 & 0.615 & -0.208 & -0.051 & 0.54 \\
\hline
LSST u & 3751.20 & -0.907 & -0.464 & -0.018 & 1.47 \\
LSST g & 4740.66 & -0.520 & -0.358 & -0.030 & 1.05 \\
LSST r & 6172.34 & 0.036 & -0.295 & -0.050 & 0.77 \\
LSST i & 7500.97 & 0.492 & -0.229 & -0.051 & 0.58 \\
LSST z & 8678.90 & 0.888 & -0.164 & -0.048 & 0.47 \\
LSST y & 9711.82 & 1.254 & -0.098 & -0.042 & 0.39 \\
\hline
Roman F062 & 6141.54 & 0.038 & -0.289 & -0.046 & 0.68 \\
Roman F087 & 8650.97 & 0.887 & -0.162 & -0.047 & 0.44 \\
Roman F106 & 10465.04 & 1.594 & -0.073 & -0.043 & 0.32 \\
Roman F129 & 12759.99 & 2.267 & -0.081 & -0.052 & 0.24 \\
Roman F146 & 13049.63 & 2.409 & -0.094 & -0.053 & 0.18 \\
Roman F158 & 15577.83 & 2.816 & -0.146 & -0.066 & 0.17 \\
Roman F184 & 18290.96 & 3.156 & -0.151 & -0.066 & 0.14 \\
Roman F213 & 21116.87 & 3.525 & -0.118 & -0.063 & 0.11 \\
\hline
2MASS J & 12350.00 & 2.200 & -0.062 & -0.049 & 0.26 \\
2MASS H & 16620.00 & 3.015 & -0.153 & -0.067 & 0.16 \\
2MASS Ks & 21590.00 & 3.589 & -0.108 & -0.062 & 0.11 \\
\hline
WISE W1 & 33526.00 & 4.377 & -0.237 & -0.092 & 0.06 \\
WISE W2 & 46028.00 & 4.586 & -0.627 & -0.153 & 0.04 \\
\enddata
\end{deluxetable}

\section{Computing Accretion Luminosity} \label{sec:Applications}

Arriving at a luminosity estimate (in our case the $L_\mathrm{acc}$) from photometric data requires two major corrections prior to applying the bolometric corrections presented in Section \ref{sec:bolCorr}. The first is to determine the extinction to the source in the filters being used. The second is to correct for distance to the source by computing the distance modulus. 

The unique SED and anisotropic emission from a viscous accretion disk requires that these corrections be treated differently from their analogs in standard stellar literature. We present the appropriate extinction corrections and distance modulus for FU Ori objects in the following section. We then describe how to apply the bolometric corrections to estimate $L_\mathrm{acc}$.

\subsection{$A_\lambda$ versus $A_V$ for an accretion disk SED}

An important step in the process of computing bolometric magnitudes -- before applying the bolometric corrections -- 
is the extinction correction to observed magnitudes, $m_\lambda$. When a broadband spectrum of an outbursting FU Ori system of obtainable, the $A_V$ is often estimated by scaling the spectrum to match that of another modestly reddened, well-known FU Ori object \citep{connelley_near-infrared_2018}. Another technique is to use interstellar absorption of Diffuse Interstellar Bands (DIBs), which are not otherwise present in the spectra of FU Ori objects \citep{CarvalhoHillenbrand2022}. 

Whatever the means of estimating $A_V$, applying the extinction correction to photometry in other bands requires knowing how the extincted flux in that band relates to the extincted flux in the $V$ band. 
For broadband filters like those in Gaia, the assumed SED of the source can affect the conversion between $A_V$ and the extinction in the desired filter, $A_{\lambda, \mathrm{disk}}$ \citep[as is known to be the case for stars, e.g.,][]{Zhang_colorCorrectionsForTemperatureAndExtinctions_2023ApJS}. 

To account for extinction effects, we thus also compute a suite of models with fixed $\dot{M} = 10^{-4.5} \ M_\odot$ yr$^{-1}$ and apply a range of $A_V$ values from 0 mag to 20 mag using the \citet{cardelli_relationship_1989} curve, to determine the conversion between $A_V$ and $A_{\lambda,\mathrm{disk}}$ for each filter. We then fit $A_{\lambda,\mathrm{disk}}$ versus $A_V$ with a linear model, which is applicable for $0.5 < A_V < 20$ in the bluest bands and $0.1 < A_V < 20$ for $\lambda > 1.0 \ \mu$m. The vertical offset in the linear fits is almost exactly 0 mag for all bands, so we simply give the slope, or $A_{\lambda,\mathrm{disk}}/A_V$ in Table \ref{tab:BC_tab}. For targets with $A_V > 15$, the bolometric corrections in the bluer bands may not be reliable due to the radically steepened SED.

\subsection{Distance Modulus for an Accretion Disk}
Once the extinction to an FU Ori object is known, the distance to the target should be accounted for using the distance modulus. One important detail is that the emission is not isotropic but rather from a disk. Consequently, distance modulus should be modified to account for the projected area of the inclined disk. Whereas for a given flux $F$, the luminosity of an isotropically emitting target at a distance $d$ is calculated by $L = 4 \pi d^2 F$, for a disk with inclination $i$ the expression becomes $L = 2 \pi d^2 F / \cos i$.

To compare observed disks to our model, the distance modulus, is given by
\begin{equation} \label{eq:distModulus}
    m_\lambda-M_{\lambda,disk} = 5 \log_{10}(d) - 5 + A_{\lambda, \mathrm{disk}} + C_i,
\end{equation}
where $A_{\lambda,\mathrm{disk}}$ is the extinction in the observed band, $C_i = - 2.5 \log_{10}(\cos i)$, and $i$ is the inclination of the observed disk. For a face-on system, $C_i = 0$, while for $i = 80^\circ$, $C_i = 1.901$. Assuming an isotropic distribution of inclinations, a typical value for $\cos i = 1/2$.  Thus, in the absence of information about the inclination of an object, a reasonable estimate is $C_i = 0.75$. 

\subsection{Disk Accretion luminosity}\label{sec:computeLum}

With an observed brightness in a given band, $m_\lambda$, and the proper distance modulus, the absolute magnitude of the disk $M_\mathrm{\lambda,disk}$ is computed by subtracting the distance modulus from $m_\lambda$, as in Equation \ref{eq:distModulus}.  Then, using the coefficients in Table \ref{tab:BC_tab} and Equation \ref{eq:Bcorr}, the bolometric correction $BC_{\lambda,\mathrm{disk}}$ can be computed. 
Next, the accretion brightness in absolute magnitudes is simply given by $M_\mathrm{bol,disk} = BC_{\lambda,\mathrm{disk}} + M_{\lambda,\mathrm{disk}}$. 

Finally, converting this accretion brightness into the accretion luminosity $L_\mathrm{acc} = L_\mathrm{bol,disk}$ 
in $L_\odot$, can be done in the usual way:
\begin{equation}
    L_\mathrm{acc} = 10^{-0.4(M_\mathrm{bol,disk} - M_{\mathrm{bol}, \odot}) } L_\odot,
\end{equation}
where as before, $M_{\mathrm{bol}, \odot} = 4.74$. If the stellar parameters are known or can be estimated, then the accretion rate can be subsequently inferred from $L_\mathrm{acc} = \frac{1}{2}\frac{G M_* \dot{M}}{R_\mathrm{inner}}$.

As noted above, the bolometric corrections for FU Ori disks are largest for the bluest (u, g) and the reddest (W1, W2) filters that are discussed here, and smaller at intermediate wavelengths.  For any given outburst, 
the source may be detected from quiescence to the outburst state in one or more of these bands.  So which bands are
``best" for determining an accurate accretion luminosity?  In addition to the issue of the size of the 
bolometric correction with wavelength, there is also the issue of the size of the extinction correction with 
wavelength. A final consideration is the sensitivity of different wavelength ranges to physical parameters 
of the accretion disk, specifically the outer radius of the actively accreting disk, as discussed on the next section.  

On balance, it appears as though red optical through short near-infrared bands (r,i,z,Y,J) 
optimally measure the accretion luminosity of the outburst. While they are less sensitive to the $\dot{M}$ of the system than bluer bands, they are not as significantly affected by the $A_V$. This makes them more robust to uncertainties in the adopted $A_V$ value. The reddest NIR bands are least impacted by $A_V$ but are contaminated by emission from the passive disk in the system, which decreases their capacity to directly trace the $L_\mathrm{acc}$. The rationale for these conclusions is provided below.

\section{Application to Two Well-Studied FU Ori Systems}

In this section, we consider several practical aspects of determining FU Ori accretion disk luminosities
using bolometric corrections to photometric observations.

\subsection{Comparison with previous results on $L_{acc}$} \label{sec:comp}

The reliability of the approach outlined above for estimating the accretion luminosity of FU Ori objects 
can be demonstrated by comparing results from this procedure with our previous detailed disk modeling for two objects: HBC 722 and V960 Mon. 

Both systems have existing Gaia and WISE lightcurves. The accretion luminosity values we adopt as the ``true" values for the systems are 
$L_\mathrm{true} = 100 \ L_\odot$ for V960 Mon and $L_\mathrm{true} = 90 \ L_\odot$ for HBC 722 \citep{Carvalho_V960MonPhotometry_2023ApJ, Carvalho_HBC722_2024ApJ}. For V960 Mon, this is the luminosity measured at the peak of its outburst, whereas for HBC 722 we use the luminosity measured after 2015, when the source reached its bright plateau phase following a short-lived dip immediately after the temporary peak in early 2011. 

The comparison between the $L_\mathrm{true}$ values derived from detailed multi-wavelength modelling and $L_\mathrm{acc}$ estimated here by applying BCs to single-band photometry, is given in Table \ref{tab:comp}. In most of the bands, the estimated luminosity is within 20\% of the measured luminosity with the notable exception of the WISE bands. This can be interpreted as the approximate systematic uncertainty in the method, which dominates over the typically $< 1 \%$ photometric uncertainty for data used to compute $L_\mathrm{acc}$. The BCs are calculated for and applicable to a broad range of source luminosities, from $8 \ L_\odot$ to $700 \ L_\odot$. So long as the target, regardless of luminosity, has an FU Ori-like accretion disk, this method is applicable.

\subsection{Cautions regarding WISE photometry use to compute L$_\mathrm{bol}$} \label{sec:caution}

For FU Ori systems, photometry redward of $K$ band has been demonstrated to increasingly (with wavelength) trace the passive disk component of the system \citep{Liu_fuorParameterSpace_2022ApJ, Carvalho_V960MonPhotometry_2023ApJ}.  It can also be sensitive to the outer boundary of the viscously heated region of the disk \citep{kospal_hbc722_2016A&A, Carvalho_HBC722_2024ApJ}. As a result, the WISE $W1$ and $W2$ photometry may overestimate the $L_\mathrm{acc}$ in some systems. This can be seen in the relatively large values of $L_\mathrm{acc}$ derived from $W1$ and $W2$ compared to those at other wavelengths. 

For V960 Mon, Table \ref{tab:comp} shows that $L_\mathrm{acc}(W1) \sim 2 L_\mathrm{true}$ and $L_\mathrm{acc}(W2) \sim 3 L_\mathrm{true}$. 
Consistent with these bolometric luminosity results, in the SED of V960 Mon, the viscous-disk-only model of \citet{Carvalho_V960MonPhotometry_2023ApJ} underestimates the $W1$ and $W2$ fluxes by factors of $\sim 2$ and $\sim 3$, respectively. The excess flux is accounted for in \citet{Carvalho_V960MonPhotometry_2023ApJ} by including a dust disk component in the system, beyond the active disk region, that is reprocessing the accretion luminosity. 

We find that for reasonable assumptions of $R_\mathrm{outer}$ in the viscous disk model (e.g., $R_\mathrm{outer} > 25 \ R_\odot$), the case of V960 Mon represents an extreme instance of overestimating $L_\mathrm{acc}$ from $W1$ or $W2$. For targets where the viscous disk extends farther than $R_\mathrm{outer} \sim 100 \ R_\odot$, the passive disk contribution at $3.5-5$ $\mu$m  is negligible. This is exemplified in HBC 722, where the $L_\mathrm{acc}$ using the $W2$ photometry is relatively consistent with both $L_\mathrm{true}$ and the near-infrared estimates, though larger than the visible range estimates.

\begin{deluxetable}{c|c|c}
	\tablecaption{Comparison between the measured accretion luminosities ($L_\mathrm{true}$) for HBC 722 and V960 Mon, and those we estimate using our BCs ($L_\mathrm{bol,disk}=L_\mathrm{acc}$). 
 \label{tab:comp}}
	\tablewidth{0pt}
	\tablehead{
            \colhead{} & \multicolumn{2}{c}{$L_\mathrm{acc}/L_\odot$} \\
            \hline
	    \colhead{Filter} & \colhead{V960 Mon} & \colhead{HBC 722}
	}
\startdata
        &$L_\mathrm{true}$ = $100 \pm 10$ & $L_\mathrm{true}$ = $90 \pm 9$ \\
Gaia BP & 123 & 77 \\
Gaia G & 101 &  73  \\
Gaia RP & 105  & 87  \\
\hline
ZTF g & $\cdots$ & 85  \\
ZTF r & $\cdots$ & 84  \\
\hline
2MASS J\tablenotemark{a} & 88 &  113  \\
2MASS H\tablenotemark{a} & 67 &  106  \\
2MASS Ks\tablenotemark{a} & 56 & 116  \\
\hline
WISE W1\tablenotemark{b} & 240 & 157  \\
WISE W2\tablenotemark{b} & 295 & 111  \\
\enddata

    \tablenotetext{a}{The flux in $J$, $H$, and $Ks$ is sensitive to the $R_\mathrm{outer}$ value in the model and for $R_\mathrm{outer} < 100 \ R_\odot$, $H$ and $Ks$ underestimate the $L_\mathrm{acc}$. }
    \tablenotetext{b}{The flux from both WISE bands includes significant contribution from the passive disk and may overestimate $L_\mathrm{acc}$.}
\end{deluxetable}

\subsection{Estimating disk $R_\mathrm{outer}$ from infrared colors}

In Section \ref{sec:caution}, we discussed how estimates of $L_\mathrm{acc}$ are susceptible to contamination in the $W1$ and $W2$ photometry from the non-active region of the disk.  Moving to shorter wavelength photometry, such as JHK, we identify another
consideration in estimating accurate $L_\mathrm{acc}$ values.

\citet{carvalho_V960MonSpectra_2023ApJ} demonstrated that the 1.5 to 2.2 $\mu$m region of an FU Ori object SED is sensitive to the $R_\mathrm{outer}$ value of the active disk region. This is especially the case for $25 < R_\mathrm{outer}/R_\odot < 100$, 
with larger values of $R_\mathrm{outer}$ best discriminated in the 3 to 5 $\mu$m range of the SED \citep{Carvalho_HBC722_2024ApJ}. 
This sensitivity to $R_\mathrm{outer}$ may present a challenge for interpreting $L_\mathrm{acc}$ from $K$ or $Ks$ band photometry. 

In Table \ref{tab:comp}, although the $J$-band-inferred $L_\mathrm{est}$ is nearly $L_\mathrm{meas}$, the $L_\mathrm{est}$ values for V960 Mon steadily decrease toward the redder 2MASS bands. This is consistent with the small $R_\mathrm{outer} = 25-35 \ R_\odot$ reported for the system. On the other hand, the $L_\mathrm{est}$ values for HBC 722 do not show this steady decrease with increasing wavelength and are in better agreement with the $L_\mathrm{est}$ from the visible (if slightly over-estimated). We expect this for HBC 722 because for the epoch we show here, $R_\mathrm{outer} > 100 \ R_\odot$ \citep{Carvalho_HBC722_2024ApJ}. 

\begin{figure}
    \centering
    \includegraphics[width = \linewidth]{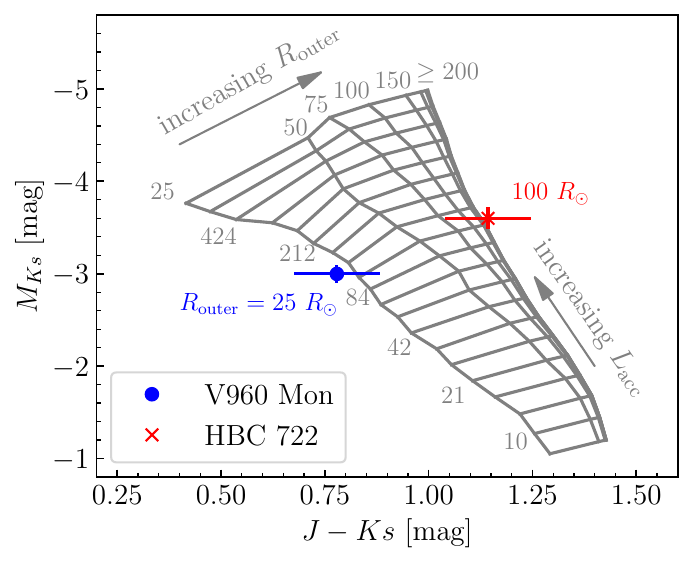}
    \caption{The $M_{Ks}$ versus extinction-corrected $J-Ks$ color magnitude space, with the locations of V960 Mon (blue point) and HBC 722 (green ``x") are marked. The grid of accretion disk model values is plotted in grey, with constant $R_\mathrm{outer}$ tracing approximately vertical lines and constant $L_\mathrm{acc}$ approximately horizontal. The values at the top of the grid mark $R_\mathrm{outer}$ in units of $R_\odot$ for each set of models and values along the lower boundary of the grid mark a few $L_\mathrm{acc}$ values in units of $L_\odot$ for reference. Models with $R_\mathrm{outer} > 200 \ R_\odot$ converge in this color space. Notice that both V960 Mon and HBC 722 lie along tracks that agree with their measured $R_\mathrm{outer}$ values (blue and red text, respectively) in this space.
    }
    \label{fig:JminusKs}
\end{figure}

Assuming that the $J$ band is a reasonable estimate of $L_\mathrm{acc}$ for $R_\mathrm{outer} > 25 \ R_\odot$, we can use the $M_J-M_{Ks}$ color of the system to estimate $R_\mathrm{outer}$. We demonstrate this in Figure \ref{fig:JminusKs}. Although the $M_{W1}-M_{W2}$ space would in principle be more sensitive to the $R_\mathrm{outer}$ in a system, the potentially significant contribution of the passive disk can confound the diagnostic. In the $M_{Ks}$ versus $M_J-M_{Ks}$ color magnitude space, the models computed for different $R_\mathrm{outer}$ are well-separated while remaining minimally affected by passive disk emission.


\section{Conclusions} \label{sec:conclusions}

Using our semi-empirical viscous disk model, we have computed bolometric corrections between observed and absolute magnitudes 
for YSOs that are dominated by luminosity generated in accretion disks. We have also provided a guide for the full set of considerations needed in order to convert a brightness measurement $m_\lambda$ for an FU Ori accretion disk into an $L_\mathrm{acc}$ estimate, using bolometric corrections. 

Our procedure accounts for the ways in which computing $L_\mathrm{acc}$ for a disk differs from the canonical methods developed to compute $L_\mathrm{bol}$ for stars, particularly due to the unique shape of the disk SED and the projection of flux in inclined disks. The shape of the SED impacts both the bolometric correction and the conversion from $A_V$ to $A_\mathrm{\lambda,disk}$ because the flux is distributed across a much broader wavelength range than in stellar SEDs. Larger disk inclinations lead to a smaller fraction of $L_\mathrm{acc}$ reaching an observer, 
which can lead to systematic underestimation of $L_\mathrm{acc}$ if the emission is assumed to be isotropic. We also add a correction term to the traditional distance modulus calculation to incorporate the inclination of the disk when computing the absolute magnitude of accretion disks. 

When applying the bolometric corrections we calculated to a typical FU Ori disk, the bands that will give bolometric-correction-derived luminosities most consistent with the true $L_\mathrm{acc}$ are those between the red-optical and $J$ (or $H$) band. This wavelength range is where the SED is flattest and least affected by the assumed $A_V$ or the presence of hot dust emission in the system. 



Estimating $L_\mathrm{acc}$ with the bolometric corrections presented here will enable a self-consistent means of comparing accretion luminosities of FU Ori objects for several ongoing and upcoming all-sky-surveys. This will facilitate building a sample of objects that can be studied uniformly as more outbursts are detected, leading to statistical robustness for population studies. 


Although we have computed the quantities and relations in the paper with FU Ori accretion disks in mind, there are elements of the work that can be applied to other disk systems. One that is universal is the need to correct for flux projection when estimating a disk luminosity from photometric data. The modified distance modulus we present in Equation \ref{eq:distModulus} is applicable to calculating the absolute magnitude of emission from any astrophysical disk. The bolometric corrections are generally applicable for viscous accretion disks with maximum temperatures ranging from 3,300 K to 10,500 K. However, due to the modification to the \citet{Shakura_sunyaev_alpha_1973A&A} temperature profile, the results will be most reliable for redder bands, as bluer bands are more sensitive to the details of $T_\mathrm{eff}(r < \frac{49}{36}R_\mathrm{inner})$.

\section{Acknowledgements}
The authors thank Antonio Rodriguez for insightful conversations and suggestions. 

\bibliography{references}{}
\bibliographystyle{aasjournal}



\end{document}